\documentclass[preprint,aps,nofootinbib,tightenlines]{revtex4}

\begin{document}

\vspace*{1.8cm}
\title{Measuring the Shape of the Extra Dimension}

\author{\vspace*{0.3cm} Ian Low}
\vspace*{.6cm}
\affiliation{
\vspace*{0.3cm}
School of Natural Sciences, Institute for Advanced Study, Princeton, NJ 08540
\vspace*{0.5cm}}

\begin{abstract}
\vspace*{0.5cm}
We study the possibility of extracting geometric information on the shape of the extra
dimension from four-dimensional data such as the mass of the Kaluza-Klein (KK)
mode. Assuming one compact extra dimension whose geometry can be considered as perturbations
in the flat background,
we show that if there is a $Z_2$ symmetry in the extra dimension, for example the KK parity
in models with Universal Extra Dimensions, then the
warp factor in the metric is completely determined by the KK mass alone.
Without KK parity,
additional information depending on the boundary conditions
 is needed to fully reconstruct the metric, even though such information 
may be experimentally challenging to obtain. The case in a general
background geometry is also considered.

\end{abstract}

%\pacs{}

\maketitle

\section{Introduction}
\label{sec:introduction}

Theories with extra dimensions have attracted enormous attention in particle physics
in the last decade. They not only provide new avenues for theoretical 
explorations, but also offer the exciting prospect of playing an active role in the upcoming
collider experiments. Starting with the revelation that extra dimensions could be
as large as the submillimeter distance and the scale of quantum gravity could be at TeV
\cite{Arkani-Hamed:1998rs,Antoniadis:1998ig,Arkani-Hamed:1998nn}, it was realized
that warped extra dimensions \cite{Randall:1999ee,Randall:1999vf} could have novel
features to address issues ranging from electroweak to gravitational physics. Ever since
there has been an explosion in the number of extra dimensional models inspired by either
the large or warped extra dimensions.

From the four-dimensional (4D) perspective extra dimensions, if there, manifest themselves through a series
of KK modes for every particle that propagates in the bulk. Typically
one starts with a given metric in the extra dimensions, assumes some boundary conditions for the bulk
fields, and then computes the KK masses from the metric. Sometimes it is assumed that all the standard
model (SM) fields are confined on a three-brane, in which case only the graviton would have a massive 
spin-2 KK tower, while sometimes all or part of the SM could live in the extra dimensions. In 
most cases the size of the extra dimensions is at TeV scale or higher, suggesting the first KK mass 
in the TeV order as well. If this is the case, it seems that the second or higher KK mode might lie beyond the
reach of the Large Hadron Collider (LHC). 
An interesting exception is the Universal Extra Dimensions (UEDs) \cite{Appelquist:2000nn, Cheng:2002ab}, 
in which not only
all the SM particles propagate in the compactified bulk, giving the hope that one may be able to 
observe another copy of
SM in the first KK level, but also the lower bound on the size of the extra dimensions is only 
at 300 GeV or
so, raising the attractive 
possibility that more than one KK level can be discovered at the LHC.

On a separate front, given the imminent start of the LHC, there are recently strong interests in the
inverse problem of interpreting the underlying physics from LHC data \cite{Arkani-Hamed:2005px}.
The goal is to study the map from the signature space of LHC to the parameter space of theoretical
models. In the context of extra dimensional models, the traditional forward approach is to study 
phenomenological consequences of a given model in a particular spacetime background, such as computing
the KK masses from the postulated metric. In this paper we consider the 
LHC inverse problem in the extra dimensional context. We will focus on the most
obvious 4D observable, the KK masses, and ask how one can extract geometrical properties of the
extra dimensions. We are interested in questions like, if one assumes a compactified extra dimension, 
what can we learn from the mass of the first KK mode? What about the second KK mass? Even if we knew
all the KK masses, would the shape of the extra dimension be uniquely determined?

It is our purpose to study the aforementioned questions in this paper,
which is organized as follows. In section II we set the stage by 
considering a five-dimensional $U(1)$ Yang-Mills
theory on a finite interval, as well as its KK decompositions. By transforming
the eigenvalue equation into a Schr\"odinger equation, 
in section III we use the time-independent
perturbation theory, as well as the reflection symmetry of the background geometry,
to study the inverse problem assuming the Neumann boundary
conditions (BCs). In section IV we consider a general background geometry
without assuming any symmetry property. The treatment here closely follows
the mathematical technique of solving the Dirichlet 
inverse eigenvalue problem for flat background in Ref.~\cite{ist}. The approach in this section 
also allows us to extend the study to other types of BCs, which
is done in section V.
Then in section VI we conclude with some discussion.

\section{$U(1)$ gauge theory on an interval}
We start with assuming 4D Lorentz invariance and one finite extra dimension which can be thought of
as an interval. Without loss of generality, the metric can be written in warped form 
\begin{equation}
ds^2 = g_{MN}\ dx^M dx^N= e^{2A(z)}\left( \eta_{\mu\nu} dx^\mu dx^\nu  - dz^2\right) , \quad 0 \le z \le L,
\end{equation}
where $\eta_{\mu\nu}={\rm diag}(1,-1,-1,-1)$ and we call
$A(z)$ the warp factor. We also use the convention that the capital Roman letters $M,N=0,1,2,3,z$
are contracted with $g_{MN}$ whereas the Greek letters $\mu,\nu=0,1,2,3$ are contracted with
$\eta_{\mu\nu}$.
When the warp factor is constant, the resulting space is
a flat extra dimension compactified on a circle, $z \simeq z + 2L$, with the projection,
$z \simeq -z$,
which is the $S^1/Z_2$ orbifold. Because of the orbifold projection all the fields living in the bulk
must be either even or odd under $z \to -z$. Even (odd) fields have Neumann
(Dirichlet) BCs at the boundaries $z=0, L$.
Usually this is the main
motivation for considering an orbifold compactification, because the
zero modes for the odd fields are projected out, which is  crucial in terms of getting
a chiral zero mode for the fermion if the SM is to live in the bulk. We wish to consider the possibility
that the warp factor is non-trivial.

As an illustration let us consider an abelian vector 
field $A_M(x,z)$ propagating on an interval $0 \le z \le L$ \cite{Davoudiasl:1999tf,Muck:2001yv}:
\begin{eqnarray}
S &=& \int \sqrt{g}\, d^4x dz\, \frac{-1}{4g_5^2} F_{MN} F^{MN} \\
  &=& \int d^4x \int_0^L dz\, \frac{e^{A(z)}}{g_5^2}\, \left( -\frac14 F_{\mu\nu} F^{\mu\nu} + \frac{1}2
        (\partial_z A_\mu - \partial_\mu A_z)^2 \right).
\end{eqnarray}
In this paper we will only be concerned with classical physics and neglect the gauge-fixing
and ghost terms.
We will also choose a gauge where $A_z = 0$ and the Neumann BCs for $A_\mu$
\begin{equation}
A_\mu^\prime(x,0) = A^\prime_\mu(x,L)=0,
\end{equation}
where $^\prime$ denotes $\partial_z$. The Neumann BCs are normally chosen to ensure a massless
zero mode. However, there could be other types of BCs if there are bulk mass terms or 
scalars at the boundaries of the interval \cite{Csaki:2003dt}, so later we will generalize the results
to Dirichlet as well mixed BCs.
Performing the KK expansion
\begin{equation}
A_\mu(x,z) = \sum_n A_\mu^n(x) f_n(z),
\end{equation}
the action becomes diagonal in the KK basis
\begin{equation}
S= \int d^4x \sum_n \left( -\frac14 F_{\mu\nu}^n F^{n\, \mu\nu} + \frac12 m_n^2 A_\mu^n A^{n\, \mu}
      \right),
\end{equation}
if the KK profiles $f_n(z)$ satisfy the equations
\begin{eqnarray}
\label{eom1}
&& \partial_z \left( e^{A(z)} \partial_z f_n \right) + m_n^2 e^{A(z)} f_n = 0, \quad
   f_n^\prime(0) = f_n^\prime(L) = 0\,; \\
&& \frac1{g_5^2} \int_0^L dz \, e^{A(z)}f_n f_m = \delta_{mn}.
\end{eqnarray}
From now on we will ignore the $1/g_5^2$ factor in the orthogonality condition as it is irrelevant to
our analysis.
Eq.~(\ref{eom1}) is an equation of Sturm-Liouville type with Neumann BCs. The 
question we are interested is essentially the inverse eigenvalue problem of the above equation:
given the mass eigenvalues, what can we learn about the warp factor $A(z)$?

To proceed, it is convenient to transform Eq.~(\ref{eom1}) into a non-relativistic Schr\"odinger 
equation \cite{Randall:1999vf}:
\begin{eqnarray}
\label{eig0}
&& f_n(z) = e^{-A(z)/2} \psi_n(z), \\
&& -\psi_n^{''} + V(z) \psi_n = m_n^2 \psi_n \, , \\
\label{kkp0}
&& V(z) = \frac12 A^{\prime\prime} + \frac14 ( A^\prime)^2,
\end{eqnarray}
We will call
$V(z)$ the KK potential associated with the warp factor $A(z)$.
It is worth noting that in Eq.~(\ref{eig0}) the BCs of the original KK wave functions $f_n(z)$ do not 
translate simply into the BCs of the new $\psi_n(z)$; the boundary values of $A(z)$ are involved as well.
For example, Neumann BCs for
the $f_n$ translate into Neumann BCs for the $\psi_n$ only if one further assumes
$A^\prime(0) = A^\prime(L)=0$. On the other hand, in solving for the warp factor from a given
KK potential in Eq.~(\ref{kkp0}), the assumed BCs for the warp factor would presumably give a unique solution.
For example, in the flat space case, $V=0$, the BCs $A^\prime(0) = A^\prime(L)=0$
give a unique, albeit trivial, answer $A(z)=0$. Otherwise the general solution for $V(z)=0$ in 
Eq.~(\ref{kkp0}) looks like $A(z)=c_1 + 2 \log(z+c_2)$, where $c_1$ and $c_2$ are integration constants.
Therefore from now on we will assume suitable BCs for $A(z)$ so that 
the BCs translate in Eq.~(\ref{eig0}).

\section{Neumann Inverse problem}

In this section we study the Neumann inverse spectral problem of the Schr\"odinger equation
\begin{equation}
-\psi^{\prime\prime} + V \psi = \lambda \psi,\quad \psi^\prime(0)=\psi^\prime(L) = 0.
\end{equation}
The idea is that when the KK potential $V(z)$ can be considered as perturbations on the flat
background, $V=0$, we can use time-independent perturbation theory of the Schr\"odinger 
equation.

In the flat space limit, the unperturbed solutions are 
\begin{equation}
\label{eig00}
\lambda_n^{(0)} = \frac{n^2\pi^2}{L^2}, 
   \quad \psi_n^{(0)} = \sqrt{\frac2{L}} \cos\left(\frac{n\pi}{L}z\right),
\end{equation}
from which we see there is a massless zero mode with constant wave function.
To the first order in perturbation, the KK masses and wave functions are,
for $n>1$,
\begin{eqnarray}
\label{eig2}
\lambda^{(1)}_n &=& \int_0^L dz \left[\psi_n^{(0)}(z) \right]^2 V(z) \nonumber \\
     &=& \frac1{L} \int_0^L dz\,V(z) + \frac1{L} \int_0^L dz \cos\left(\frac{2n\pi}{L}z\right) V(z), \\
\label{eigf0}
\psi_n^{(1)}(z) &=& \sum_{m \neq n} \frac{\psi_m^{(0)}(z)}{\lambda^{(0)}_n - \lambda^{(0)}_m}
                 \int_0^L dt\, \psi_n^{(0)}(t) \psi_m^{(0)}(t) V(t). 
\end{eqnarray}
The zero mode $n=0$ is a special case and needs to be singled out from perturbation because
its masslessness is guaranteed by the 4D gauge invariance.\footnote{I am grateful
to Yuri Shirman and Arvind Rajaraman for bringing this issue to my attention.}
Indeed, the constant wavefuction is always a solution with zero eigenmass in 
Eq.~(\ref{eom1}), which implies the exact zero mode wavefunction $\psi_0(z) = \exp(A/2)$.
On the other hand, if we had chosen differen BCs, there would have been no massless zero mode
and no need to single it out in perturbation.
From Eq.~(\ref{eig2}) we immediately see that the first-order corrections
to flat-space KK masses are related to coefficients of the Fourier cosine
series of the KK potential. In particular, the correction to the $n$th eigenmass
is related to the sum of the average of the KK potential and the $n$th coefficient
of Fourier cosine series. Therefore,
higher KK masses probe the metric at shorter distances, in accordance with usual intuition.

One important observation following from Eq.~(\ref{eig2}) is the fact that the KK masses
are only sensitive to the even part of the KK potential with
respect to reflections on the mid-point of the interval $z \to L-z$. Unless this $Z_2$ reflection
is a symmetry of the extra dimension, KK masses alone are not sufficient to uniquely determine
the KK potential. Nevertheless, 
such a geometric $Z_2$ reflection is none other than the KK parity
in UEDs \cite{Appelquist:2000nn,Cheng:2002ab}. In UEDs with one extra dimension the SM propagate
in 5D compactified on the orbifold $S^1/Z_2$. For theories compactified on a circle $S^1$, momentum
conservation in the 5th direction implies conservation of the KK number at each interaction vertex.
When considering $S^1/Z_2$, however, the orbifold has fixed points at the boundaries which break
the translational invariance, and hence momentum conservation, in the 5th direction. Moreover,
quantum corrections in the bulk induces divergent terms on the two boundaries that renormalize
localized 4D interactions there \cite{Georgi:2000ks,Cheng:2002iz}. In the end only a $Z_2$ subgroup of the
translational invariance, that is reflections with respect to the mid-point $z=L/2$, is preserved,
which is called KK parity \cite{Cheng:2002ab}. For phenomenological considerations, 
the KK parity is defined as a flip of the line interval about the center $z=L/2$ combined with a
$Z_2$ transformation that changes the sign of all fields odd under the orbifold projection, which
are fields that have Dirichlet BCs. This is so that all the even number
KK modes are invariant, while the odd number KK modes change sign, under the KK parity. Our finding
is that for extra dimensional models that have the KK parity, the shape of the extra dimension is
completely determined by measurements of KK masses.

A new $Z_2$ parity for theories beyond SM, under which the SM is even and 
(some of) the new particles are odd, is in fact very well-motivated phenomenologically. 
Perhaps the most
prominent feature
of such a $Z_2$ parity is suppressions of precision electroweak contributions from the new particles
\cite{Cheng:2003ju},
rendering their masses light at or below 1 TeV and allowing for a solution to the little hierarchy
problem. Another important feature is the existence a stable particle, that is the
lightest particle charged under the parity, which is a good candidate for dark matter if it is 
electrically neutral.
Examples of such a parity, other than the KK parity, are the $R$ parity in supersymmetry and the $T$ parity
in little Higgs models \cite{Cheng:2003ju,Cheng:2004yc,Low:2004xc}.

Without KK parity, it is 
natural to ask is how to determine the coefficients of the Fourier sine series of the KK potential from
four-dimensional data. To this end we notice that the coefficient of the Fourier sine series
\begin{equation}
\label{sine0}
s_n=\frac{2}L \int_0^L dz\, V(z) \sin \left(\frac{2n\pi z}{L} \right)
\end{equation}
only depends on the KK odd part of $V(z)$, and as such is a measure of the breaking of the KK parity.
Therefore any quantity that is sensitive to violations of KK parity will be related to the Fourier
sine coefficients. One such quantity is the absolute value of the ratio $|\psi_n(z)/\psi_n(L-z)|$.
If KK parity is a good symmetry and the geometry is symmetric with respect to reflections about
the mid-point of the interval, then the ratio should be unity. The above argument suggests
the definition 
\begin{equation}
\label{kap0}
\kappa_n(z) = \log \left|\frac{\psi_n(z)}{\psi_n(L-z)}\right|,
\end{equation}
which vanishes when KK parity is conserved. Using Eqs.~(\ref{eig00}) and (\ref{eigf0}),
we can derive an expression for $\kappa_n(z)$ in perturbation:
\begin{eqnarray}
\label{kap1}
\kappa_n(z) &=& \sum_{m\neq n} \frac{1-(-1)^{m-n}}{\lambda_n^{(0)}-\lambda_m^{(0)}}
           \ \frac{\psi_m^{(0)}(z)}{\psi_n^{(0)}(z)} \int_0^L dt\, \psi_m^{(0)}(t) \psi_n^{(0)}(t) V(t)
                \\
\label{kap2}
 &=& \sum_{m\neq n}  \frac{1-(-1)^{m-n}}{(n^2-m^2)(\pi^2/L^2)}
          \  \frac{\cos \frac{m \pi z}{L}}{\cos \frac{n \pi z}{L}} 
         \frac2L \int_0^L dt\, \cos\left(\frac{m\pi t}L \right) \cos\left(\frac{n\pi t}L \right) V(t).
\end{eqnarray}
In the above, because of the coefficient $1-(-1)^{m-n}$, the summation effectively only runs over 
those $m$'s for which $m+n$ is an odd integer. For this case, the cosines in the  integrand
has odd parity under $z\to L-z$ and the integral is non-vanishing only if $V(t)$ has a KK odd component.
In principle, one could work out the Fourier sine series of the integrand in Eq.~(\ref{kap2})
\begin{equation}
 \cos\left(\frac{m\pi t}L \right) \cos\left(\frac{n\pi t}L \right) = \sqrt{\frac2L}\sum_k a_k 
     \ \sin \frac{2\pi k t}L,
\end{equation}
 from which a relation between $\kappa_n(z)$ and the $s_n$ in Eq.~(\ref{sine0}) follows. However,
following a suggestion for a similar quantity for the Dirichlet inverse problem in \cite{ist}, one
can show that $\kappa_n(0)$ is directly proportional to the Fourier sine coefficients $s_n$,
\begin{eqnarray}
\label{kap3}
\kappa_n(0) &=& \sum_{m\neq n}  \frac{1-(-1)^{m-n}}{(n^2-m^2)(\pi^2/L^2)}
          \   
         \frac2L \int_0^L dt\, \cos\left(\frac{m\pi t}L \right) \cos\left(\frac{n\pi t}L \right) V(t)
   \\
&=& \frac{L}{2\pi n} \, \int_0^L dz\, V(z) \sin \left(\frac{2n\pi z}{L} \right),
\end{eqnarray}
if one uses the identity
\begin{equation}
\label{idd0}
\frac{L}{n\pi} \sin \frac{n\pi z}L = 
 \sum_{m\neq n} \frac2L \cos \frac{m \pi z}L\frac{1-(-1)^{m-n}}{(n^2-m^2)(\pi^2/L^2)}, \quad 0 \le z \le L.
\end{equation}
One way to derive the above identity is to use the Green's function with Neumann BCs
\begin{eqnarray}
&& G(z,z') = \sum_n \frac{\psi_n^{(0)}(z)\psi_n^{(0)}(z')}{\lambda - \lambda_n^{(0)}}, \\
&& -\partial_z^2 G(z,z') - \lambda G(z,z') = \delta(z-z'), 
\end{eqnarray}
and then plug into
\begin{equation}
\sin \sqrt{\lambda}z' = \int_0^L dz\ \delta(z-z') \sin \sqrt{\lambda} z,
\end{equation}
which is just the expansion of the sine
function in the complete basis $\{\psi_n^{(0)}(z), 0 \le z \le L\}$. 

In terms of the original eigenfunctions in Eq.~(\ref{eig0}), 
\begin{equation}
\log \left| \frac{f_n(0)}{f_n(L)} \right| = -\frac12(A(0)-A(L))
+ \frac{L}{2\pi n} \, \int_0^L dz\, V(z) \sin \left(\frac{2n\pi z}{L} \right), \quad n>0.
\end{equation}
That is $\kappa_n(0)$  measures the difference in the
boundary values of the warp factor $A(z)$, as well as 
the Fourier sine coefficients of the KK potential. Unfortunately, it appears that the ratio
of the boundary values of the wave functions is not easily accessible from the experimental
perspective; one needs to be able to resolve the extra dimension and make a comparison at two
opposite points. What is worse,
 as mentioned earlier quantum corrections in the bulk will induce
logarithmically divergent contributions to the gauge kinetic terms that are localized on the 
boundaries \cite{Georgi:2000ks,Cheng:2002iz}. Thus from the viewpoint of 4D effective field theories, 
the values of the wave functions on the orbifold fixed points may even be arbitrary and theoretically
incalculable due to their
UV sensitivity.

On the other hand, there are certainly low-energy 
observables that probe the breaking of KK parity. Suppose
we extend the $U(1)$ gauge theory to a non-abelian theory, then there are three-point couplings
$g_{lmn}$ as well
as four-point couplings $g_{klmn}$ of different KK modes, where the indices denote the KK numbers. 
If KK parity is a good quantum number, $g_{lmn}=0$ for odd integral $l+m+n$ and $g_{klmn}=0$ 
for odd integral $k+l+m+n$. A non-zero value for either of them would indicate breaking of
KK parity and potentially probe the KK odd part of $V(z)$. 
Nevertheless, the relations between the three/four-point couplings and the Fourier
sine coefficients are contaminated by the warp factor itself. 
As an example, consider the three-point couplings
\begin{eqnarray}
\label{3pt}
g_{lmn} &\propto& \int_0^L dz\, e^{A(z)} f_l(z) f_m(z) f_n(z) \\
    &=& \int_0^L dz\, e^{-A(z)/2} \psi_l(z) \psi_m(z) \psi_n(z).
\end{eqnarray}
The product of the $\psi(z)$'s in the integrand could be computed in perturbation using $\psi^{(0)}(z)$.
It is also possible to express the product
 in the Fourier sine series, which however would involve an infinite number of terms. Unfortunately,
the warp factor also goes into the integrand. Thus
without knowing the warp factor a priori, it seems difficult, if not impossible, to actually perform
the integration and extract the desired Fourier coefficients from the three-point couplings. It is
in fact possible to eliminate the warp factor in the integrand in Eq.~(\ref{3pt}) by taking
advantage of the fact that the zero mode wavefunction is constant. For example, 
choosing $l=0$ we have
\begin{equation}
g_{0mn} \propto \int_0^L dz\, \psi_m(z) \psi_n(z)
\end{equation}
which does not involve the warp factor explicitly. Nevertheless, it is simple to check in perturbation
that the terms linear in the KK potential all cancel and only ${\cal O}(V^2)$ terms survive. Again
it is very difficult to extract the Fourier sine coefficients this way.
To sum up, the three/four-point couplings probe the KK odd part of the warp factor as well
as the KK potential, and in general it seems very difficult to disentangle these two effects
in the couplings. 
On the
other hand, if empirically it is found that all these KK odd $g_{lmn}$'s and $g_{klmn}$'s are
vanishingly small, as would be preferred from precision electroweak constraints, then one could just
use the eigenmasses to extract the Fourier cosine coefficients of the KK potential 
to reconstruct the metric.

\section{General Background}
In this section we discuss the situation when the warp factor cannot be considered as 
perturbations on flat spacetime $V(z)=0$. One example is the Anti-di Sitter
(AdS) background employed in \cite{Randall:1999ee,Randall:1999vf}, for which the metric is
\begin{equation}
ds^2 = \left(\frac{1}{k z}\right)^2 \left( \eta_{\mu\nu} dx^\mu dx^\nu - dz^2\right) ,
\end{equation}
where $k$ is the AdS curvature scale. The KK potential for the AdS background is 
\begin{equation}
V_{AdS}(z) = \frac3{4z^2} .
\end{equation}
The KK potential for the AdS space apparently does not respect KK parity, and therefore
the first KK mass is generally required to be heavier than 1 TeV or higher. Moreover,
it does not appear proper to consider the above KK potential as a perturbation on the 
flat background $V=0$ because of the singularity at $z=0$; the integral of $V_{AdS}$ 
diverges in the interval $0 \le z \le L$. On the other hand, if the warp
factor is exactly AdS, then the KK spectrum is given by roots of Bessel functions 
\cite{Pomarol:1999ad,Gherghetta:2000qt}
 and should
be identifiable. Therefore in the following we assume a KK spectrum that can be
roughly, but not exactly, identified with that coming from a known background such as
the AdS, suggesting that the real
geometry only slightly deviates from the known background and could be considered as perturbations.
Our construction in the following 
is adapted from that in \cite{ist}, which specifically considers Dirichlet
inverse spectral problem for $V(z)$ that is regular on the interval.

Assuming the warp factor in the metric to be of the form
\begin{equation}
A(z) = A_0(z) + A_1(z), \quad 0 \le z \le L,
\end{equation}
where $A_0(z)$ is a known background and $A_1(z)$ is a small fluctuation. The KK potential is then
\begin{eqnarray}
V(z) &=& V_0(z) + V_1(z) , \\
V_0 &=&  \frac12 A_0'' + \frac14 (A_0')^2 ,\\
V_1 &=&  \frac12 (A_0' A_1'+A_1'') + \frac14 (A_1')^2 .
\end{eqnarray}
That is, 
we would like to consider the Neumann inverse eigenvalue problem of the following differential equation
\begin{equation}
\label{eom2}
-y^{\prime\prime} + (V_0 + V_1) y = \lambda y,  \quad 0 \le z \le L,
\end{equation}
when $V_1$ can be considered as perturbations.
One first considers the unperturbed
equation
\begin{equation}
\label{eom20}
-y^{\prime\prime} + V_0\, y = \lambda y,
\end{equation}
and constructs the eigensystem $\{\lambda_n^{(0)}, \psi_n^{(0)}\}$
satisfying the Neumann BCs. Then as before the first-order perturbed
eigenvalues are
\begin{eqnarray}
\label{gen0}
\lambda_n &=& \lambda_n^{(0)} + \lambda_n^{(1)} \nonumber \\
   &=& \lambda_n^{(0)} + \int_0^L dz [\psi_n^{(0)}(z)]^2 V_1(z).
\end{eqnarray}
Therefore, once the first $N$ KK masses are measured, the above equation leaves
$N$ constraints on the KK potential.
In general,
the eigenvalues $\lambda_n$ and eigenfunctions $\psi_n$ are both functionally dependent
on the perturbation $V_1$. One can compute $\delta \lambda_n/\delta V_1$ by taking
the functional derivative $\delta/\delta V_1$ of the equation\footnote{Note that for the purpose of
taking functional derivative $V_0$ and $V_1$ are independent variables.}
\begin{equation}
\label{eom6}
-\psi_n'' + (V_0 + V_1) \psi_n = \lambda_n \psi_n, \quad \psi_n'(0)=\psi_n'(L)=0.
\end{equation}
Interchanging differentiations with respect to $z$ and $V_1$, multiplying both sides by $\psi_n$, and 
integrating we find
\begin{eqnarray}
&& \int_0^L dz\, \psi_n(z) \left[ -\frac{d^2}{dz^2} + (V_0+V_1)\right]
      \frac{\delta \psi_n(z)}{\delta V_1(z')} = \nonumber \\
&& \quad \quad    -[\psi_n(z')]^2 + \frac{\delta \lambda_n(V_1)}{\delta V_1(z')} +
      \int_0^L dz\, \lambda_n \psi_n(z)  \frac{\delta \psi_n(z)}{\delta V_1(z')}.
\end{eqnarray}
Since the differential equation (\ref{eom6}) is self-adjoint, we arrive at
\begin{equation}
\label{id21}
  \frac{\delta \lambda_n(V_1)}{\delta V_1(z)} = [\psi_n(z)]^2,
\end{equation}
which is indeed satisfied by Eq.~(\ref{gen0}) in perturbation. This result will be useful later.

The lesson learned from the previous section, is that
the eigenmasses only give limited information on $V_1$; in the flat space case
only the Fourier cosine coefficients are given by the eigenmasses. More information can be 
extracted by looking
at the eigenfunctions. To do so we need to consider solutions 
$Y=\{y_1,y_2\}$ of Eq.~(\ref{eom2}) satisfying
the following initial conditions
\begin{eqnarray}
\label{bc1}
y_1(0,\lambda;V_1) = y_2^\prime(0,\lambda;V_1)=1 \ ,\\
\label{bc2}
y_1^\prime(0,\lambda;V_1)=y_2(0,\lambda;V_1) = 0 \ ,
\end{eqnarray}
where we have emphasized the dependence of the solutions on $\lambda$ and $V_1$.
It is clear that every solution of Eq.~(\ref{eom2})
can be written as $y(z)=y(0) y_1(z) + y'(0) y_2(z)$. For example, with Neumann BCs we
have
\begin{equation}
\label{bc4}
\psi_n(z) = \frac{y_1(z,\lambda_n;V_1)}{||y_1(z,\lambda_n;V_1)||}, \quad
y_1(z,\lambda_n;V_1) = \frac{\psi_n(z)}{\psi_n(0)}.
\end{equation} 
where $||\cdot||$ means the norm of the function in the Hilbert space.
In addition, Eq.~(\ref{id21}) becomes
\begin{equation}
\label{id3}
\frac{\delta \lambda_n(V_1)}{\delta V_1(z)} = \frac{[y_1(z,\lambda_n;V_1)]^2}
    {||y_1(z,\lambda_n;V_1)||^2}.
\end{equation}
The fundamental solution $Y$ also has the
property that the Wronskian determinant
is unity
\begin{equation}
W(Y) = \det \left| \begin{array}{cc}
                y_1 & y_2 \\
                y_1' & y_2' 
               \end{array} \right| = 1 .
\end{equation}
This can be proven by showing that $dW/dz=0$, which follows from the fact that $Y$
satisfies Eq.~(\ref{eom2}), and using $W(0)=1$.
Then the solution of the inhomogeneous equation
\begin{equation}
\label{eom3}
-y^{\prime\prime} + (V_0+V_1) y = \lambda y - f(z)
\end{equation}
is given by
\begin{equation}
\label{eom4}
y(z, \lambda;V_1) = \int_0^z dt \left[y_1(t)y_2(z)-y_1(z)y_2(t)\right]f(t).
\end{equation}
Our objective is to show that the quantity 
\begin{equation}
\label{kap5}
\kappa_n(V_1) = -\log \left|y_1(L,\lambda_n;V_1) \right|
\end{equation}
provides additional information on $V_1$. In the flat space case, using Eq.~(\ref{bc4}), we see that
the above definition agrees with Eq.~(\ref{kap0}) and is related to the Fourier sine coefficient of
the KK potential. 

We need three identities to
complete the proof. The first one is Eq.~(\ref{id21}). For the second we need to compute
 the functional derivative of $Y=\{y_1,y_2\}$ with respect to $V_1$. Taking
$\delta/\delta V_1(z')$
 in Eq.~(\ref{eom2}) and interchanging the functional derivative with $d/dz$, we have
\begin{equation}
-\left(\frac{\delta Y(z)}{\delta V_1(z^\prime)} \right)^{\prime\prime} +
    (V_0+V_1) \frac{\delta Y(z)}{\delta V_1(z^\prime)}  = 
     \lambda \frac{\delta Y(z)}{\delta V_1(z^\prime)}  - \delta(z-z^\prime) Y(z),
\end{equation}
which is of the form in Eq.~(\ref{eom3}). Utilizing Eq.~(\ref{eom4}) we obtain
\begin{eqnarray}
\label{id1}
\frac{\delta Y(z)}{\delta V_1(z^\prime)}  &=&
   \int_0^z dt \left[y_1(t)y_2(z)-y_1(z)y_2(t)\right] \delta(z^\prime - t) Y(t) \nonumber \\
  &=& \left[y_1(z^\prime) y_2(z) - y_1(z) y_2(z^\prime) \right] Y(z^\prime)\ {\cal I}_{[0,z]}(z^\prime),
\end{eqnarray}
where the indicator function ${\cal I}$ is such that
\begin{eqnarray}
{\cal I}_{[0,z]}(z^\prime) &=& 1 \quad {\rm if }\ \ z^\prime < z, \nonumber \\
        &=& 0 \quad {\rm if} \ \ z^\prime > z.
\end{eqnarray}
From Eq.~(\ref{id1}) we could derive a similar expression for $\delta Y'(z)/\delta V_1$.
The last identity is $\partial Y / \partial \lambda$. Again differentiating Eq.~(\ref{eom2})
with respect to $\lambda$, interchanging the derivatives, and making use of Eq.~(\ref{eom4}), we have
\begin{equation}
\label{id2}
\frac{\partial Y(z)}{\partial \lambda} = 
   -\int_0^z dt \left[ y_1(t)y_2(z)-y_1(z)y_2(t) \right] Y(t).
\end{equation}
Using Eqs.(\ref{id3}, \ref{id1}, \ref{id2}), it only takes some algebra to show that
\begin{eqnarray}
\label{eom5}
\frac{\delta \kappa_n(V_1)}{\delta V_1(z^\prime)} &=& -
  \frac1{y_1(L)}\left.\left( \frac{\partial}{\partial \lambda} y_1(L,\lambda;V_1) 
            \frac{\delta \lambda}{\delta V_1(z^\prime)} +
    \frac{\delta}{\delta V_1(z^\prime)} y_1(L,\lambda;V_1)\right)\right|_{\lambda=\lambda_n} 
      \nonumber\\
  &=& y_1(z^\prime,\lambda_n;V_1) y_2(z^\prime,\lambda_n;V_1) - [\psi_n(z^\prime)]^2 
           \int_0^L dt\, y_1(t,\lambda_n;V_1) y_2(t,\lambda_n;V_1).
\end{eqnarray}
 We then have
\begin{eqnarray}
\label{kappa1}
\kappa_n(V_1) - \kappa_n(0) &=& \int_0^1 dt\, \frac{d}{dt} \kappa_n(tV_1) \nonumber \\
            &=& \int_0^1 dt \int_0^L dz \frac{\delta \kappa_n(tV_1)}{\delta (tV_1)} V_1(z) \nonumber \\
            &=& \int_0^L dz\, \frac{\delta \kappa_n(0)}{\delta V_1} V_1(z) + {\cal O}(V_1^2),
\end{eqnarray}
where we have used the definition of total derivative on a functional in the Hilbert space:
\begin{equation}
\left. \frac{d}{d\epsilon} F[q+\epsilon v] \right|_{\epsilon=0}= \int_0^L dz \frac{\delta F}{\delta q} v .
\end{equation} 
Now $\kappa_n(V)$ can be computed in perturbation:
\begin{eqnarray}
\label{kap6}
\kappa_n(V_1)-\kappa_n(0) &=& \int_0^L dz \, y_1(z,\lambda_n^{(0)};0)y_2(z,\lambda_n^{(0)};0)\, V_1(z)\nonumber \\
     \quad \quad && -\lambda_n^{(1)} 
             \int_0^L dt\, y_1(t,\lambda_n^{(0)};0)y_2(t,\lambda_n^{(0)};0) .
\end{eqnarray}
For the flat space case $V_0=0$, $\lambda_n = n^2\pi^2/L^2$ and
\begin{equation}
y_1(z,\lambda;0) = \cos \sqrt{\lambda} z, \quad y_2(z,\lambda;0) = \frac{\sin \sqrt{\lambda} z}{\sqrt{\lambda}}.
\end{equation}
Furthermore, $\kappa_n(0)=0$ and
Eq.~(\ref{kap6}) gives the Fourier sine coefficients of $V_1$.

\section{Other types of boundary conditions}

In this section we extend the results so far to other types of BCs, the Dirichlet 
and mixed BCs, with a focus on the flat space background. These BCs might be useful for 
bulk scalars or fermions on an interval.
The mixed BCs actually do not arise in orbifold compactification for its fields do not have a 
definite parity under the orbifold projection $z\to -z$. Nevertheless, if we are only concerned 
with a field theory on an interval, then it could be consistent.

The identities we derived 
so far, Eqs.~(\ref{id21}, \ref{id1}, \ref{id2}), do not depend on the BCs
we choose. However, the eigenvalues and eigenfunctions in Eq.~(\ref{eig00}), as well
as Eq.~(\ref{id3}), do depend on the Neumann BCs chosen. To generalize to other types
of BCs, the important 
observation is the following:
\begin{itemize}
\item If $ \psi_n(z) = y_1(z,\lambda_n)/||y_1(z,\lambda_n)||$,
\begin{eqnarray}
&& 
- \frac{\delta}{\delta V_1(z')} \log |y_1(L,\lambda_n;V_1)| =
- \frac{\delta}{\delta V_1(z')} \log |y_1'(L,\lambda_n;V_1)| = \nonumber \\
&& \quad \quad y_1(z^\prime,\lambda_n;V_1) y_2(z^\prime,\lambda_n;V_1) - [\psi_n(z^\prime)]^2 
           \int_0^L dt\, y_1(t,\lambda_n;V_1) y_2(t,\lambda_n;V_1) .
\end{eqnarray}
\item If $\psi_n(z) = y_2(z,\lambda_n)/||y_2(z,\lambda_n)||$,
\begin{eqnarray}
&&
- \frac{\delta}{\delta V_1(z')} \log |y_2(L,\lambda_n;V_1)| =
- \frac{\delta}{\delta V_1(z')} \log |y_2'(L,\lambda_n;V_1)| = \nonumber \\
&& \quad \quad - y_1(z^\prime,\lambda_n;V_1) y_2(z^\prime,\lambda_n;V_1) + [\psi_n(z^\prime)]^2 
           \int_0^L dt\, y_1(t,\lambda_n;V_1) y_2(t,\lambda_n;V_1) .
\end{eqnarray}
\end{itemize}
These equations can be proven along the line of proving Eq.~(\ref{eom5}).
 Now we can summarize the results for the Dirichlet as well mixed BCs in the flat background,
using the notation $\lambda_n = n^2\pi/L^2$
and $\lambda_{n+1/2} = (n+1/2)^2\pi^2/L^2$, 
\begin{itemize}
\item Dirichlet BCs $\psi(0)=\psi(L)=0$:
\begin{eqnarray}
&& m_n^2=\lambda_n , \quad
            \psi_n^{(0)}(z) = \sqrt{\frac2L} \sin \sqrt{\lambda_n} z
         = \frac{y_2(z,\lambda_n;0)}{||y_2(z,\lambda_n;0)||} ,\\
&& \lambda_n^{(1)} = \frac1{L} \int_0^L dz\, V(z) -
         \frac1{L} \int_0^L dz\, V(z)\, \cos\,  2\sqrt{\lambda_n}z , \\
&& \kappa_n(V)\equiv -\log |y_2^\prime(L,\lambda_n;V)| = -
  \log \left|\frac{\psi_n^{\prime}(L)}{\psi_n^{\prime}(0)}\right| \nonumber \\
&& \phantom{\kappa_n(V)} = \kappa_n(0)
      -\frac{L}{2\pi n} \int_0^L dz\,V(z)\, \sin\,2\sqrt{\lambda_n} z  , \\
 && \kappa_n(0)=-\log |y_2^\prime(L,\lambda_n;0)| = 0.
\end{eqnarray}

\item Mixed BCs (I) $\psi(0)=\psi^\prime(L)=0$:
\begin{eqnarray}
&& m_n^2=\lambda_{n+\frac12}, \quad
            \psi^{(0)}_n(z) = \sqrt{\frac2L} \sin \sqrt{\lambda_{n+\frac12}} z
         = \frac{y_2(z,\lambda_{n+\frac12};0)}{||y_2(z,\lambda_{n+\frac12};0)||} ,\\
&& \lambda_n^{(1)} = \frac1{L} \int_0^L dz\, V(z) -
         \frac1{L} \int_0^L dz\, V(z)\, \cos\,  2\sqrt{\lambda_{n+\frac12}}z , \\
&& \kappa_{n}(V)\equiv -\log |y_2(L,\lambda_{n+\frac12};V)| = -
  \log \left|\frac{\psi_n(L)}{\psi_n^{\prime}(0)}\right| \nonumber \\
&& \phantom{\kappa_{n}(V)} =
      \kappa_n(0)- \frac{L}{2\pi (n+1/2)} \int_0^L dz\,V(z)\, \sin\,2\sqrt{\lambda_{n+\frac12}} z  , \\
 && \kappa_{n}(0)= -\log |y_2(L,\lambda_{n+\frac12};0)| = \frac12 \log \lambda_{n+\frac12}.
\end{eqnarray}

\item Mixed BCs (II) $\psi^\prime(0)=\psi(L)=0$:
\begin{eqnarray}
&& m_n^2=\lambda_{n+\frac12}, \quad
            \psi_n^{(0)}(z) = \sqrt{\frac2L} \cos \sqrt{\lambda_{n+\frac12}} z
         = \frac{y_1(z,\lambda_{n+\frac12};0)}{||y_1(z,\lambda_{n+\frac12};0)||} ,\\
&& \lambda_{n+\frac12}^{(1)} = \frac1{L} \int_0^L dz\, V(z) +
         \frac1{L} \int_0^L dz\, V(z)\, \cos\,  2\sqrt{\lambda_{n+\frac12}}z , \\
&& \kappa_{n}(V)\equiv -\log |y_1^\prime(L,\lambda_{n+\frac12};V)| = -
  \log \left|\frac{\psi_n^\prime(L)}{\psi_n(0)}\right| \nonumber \\
&& \phantom{\kappa_{n}(V)} = \kappa_n(0)
    + \frac{L}{2\pi (n+1/2)} \int_0^L dz\,V(z)\, \sin\,2\sqrt{\lambda_{n+\frac12}} z  , \\
 && \kappa_{n}(0)=- \log | y_1^\prime(L,\lambda_{n+\frac12};0)| = -\frac12 \log \lambda_{n+\frac12}.
\end{eqnarray}

\end{itemize}

\section{Discussion and conclusion}

In this paper we studied the problem of reconstructing the metric of the extra dimension using
four-dimensional data. When the geometry can be considered as perturbations in a flat background,
we showed that the deviation of each KK mass from the exact flat space limit
 gives the Fourier cosine coefficient of a KK potential, which
is related to the warp factor through a non-linear second-order differential equation. 
If the KK parity, reflections about the mid-point of the extra dimension, is a good symmetry of the
theory, then the Fourier sine coefficient of the KK potential vanishes and the metric can be 
determined by measuring KK masses alone. On the other hand, if KK parity is not a symmetry, then
the boundary values of each wave function are necessary to determine the Fourier sine coefficient
of the KK potential. Such information, nonetheless, seems challenging to obtain experimentally for
one needs to resolve the size of the extra dimension first and then make comparison of the
wave function at two opposite ends. If there are brane localized interactions at the boundaries,
as required by quantum corrections coming from bulk fields, then it might be possible
to probe the values of the wave functions at the boundaries. However, because of the UV sensitivity
of these brane localized terms, their strength is not calculable within the low-energy effective
theory. There are averaged quantities sensitive to the breaking of KK parity such as the three-
and four-point couplings of the non-abelian gauge fields. However these couplings generally involve
many different Fourier sine coefficients of the KK potential. Moreover, they also probe
the KK odd part of the warp factor and, therefore, do not provide direct
access to Fourier coefficients of the KK potential without prior knowledge of the warp factor.

A general background geometry other than the flat space is also considered in this paper. The 
possibility arises when the KK potential of the geometry has non-integrable
singularities on the
interval and cannot be considered as perturbations in the flat background. One example is the
AdS geometry whose KK potential grows like $1/z^2$ as $z\to 0$. In this situation
three types of BCs: Neumann, Dirichlet,
and the mixed, are considered. Generically information on the behavior of the wave function at the
boundaries of the extra dimension provides constraints on the KK potential in addition to those
coming from the KK mass.

To implement the idea in this paper in the real world, one needs to first identify the background
spacetime on which the geometry can be considered fluctuations. For example whether the KK spectrum
roughly fits the flat space spectrum, which is evenly-spaced, or the AdS spectrum, which is the root 
sequence of Bessel functions. Obviously this would require measurements of several KK masses,
even though
 realistically it is not clear one would be able to measure more than one KK level, if at all,
in the near future,
as the KK mode
is generally expected to be heavier than 1 TeV from various constraints. An exception in this regard
is the UEDs, for which the compactification scale can be as low as 300 GeV, raising the prospect of
observing several KK levels. 
In UEDs this is possible because of the KK parity, a $Z_2$ reflection
about the mid-point of the extra dimension, which is strongly suggested by precision electroweak 
measurements. If KK parity is indeed a good symmetry, even for non-flat geometry, then the measurement
of $N$ KK masses could provide useful information on the first $N$ Fourier cosine coefficients of the
KK potential, if the KK spectrum fits approximately that from a flat extra dimension.
 However, because it is the deviation from $n^2\pi^2/L^2$,
 the flat space limit, that gives the sum of the average as well as the $n$th Fourier cosine
coefficient (see Eq.~(\ref{eig2})), and the size of the extra dimension $L$ is unlikely to be known a priori, in reality
one could probably only hope for an $(N+2)$-parameter fit using $N$ measured KK masses. On the other
hand,
it will be important to understand the extent of KK parity violation through KK odd 
processes like the
decay of the first KK mode into two zero modes, or inelastic scattering of two zero modes into
one first KK mode and one zero mode. These information will be an indication on the size of
the Fourier sine coefficients of the KK potential.

Another approach to the inverse problem discussed in this paper is to discretize the Sturm-Liouville
equation and turn the problem into the matrix inverse eigenvalue problem. Physically speaking this 
amounts to using deconstruction \cite{Arkani-Hamed:2001ca,Hill:2000mu} to approximate the continuous
extra dimension. However, there is some subtlety due to the mismatch of eigenmasses in the high
energy between
the deconstruction and the continuous case. Such an approach is currently under investigation
\cite{low}.

\begin{acknowledgments}
This work was supported by the Department
of Energy under grant DE-FG02-90ER40542. It is a pleasure to acknowledge
correspondence with H. C. Cheng, J. Erlich, G. Kribs, and M. Wise. This work was completed while
visiting the particle theory group at University of California at Irvine, whose hospitality is appreciated.
In addition, I am grateful to J. A. R. Cembranos, A. Rajaraman, and Y. Shirman for helpful discussion.
\end{acknowledgments}

\section*{Note Added}
After this work was completed, Ref.~\cite{Rabadan:2002wb} came to my awareness, which
also considered the problem of reconstructing the geometrical properties of a compact manifold from
KK masses, albeit with a very different approach. I would like to thank M. Kleban for bringing
this reference to my attention.


\begin{thebibliography}{nn}

%\cite{Arkani-Hamed:1998rs}
\bibitem{Arkani-Hamed:1998rs}
N.~Arkani-Hamed, S.~Dimopoulos and G.~R.~Dvali,
%``The hierarchy problem and new dimensions at a millimeter,''
Phys.\ Lett.\ B {\bf 429}, 263 (1998)
[arXiv:hep-ph/9803315].
%%CITATION = HEP-PH 9803315;%%

%\cite{Antoniadis:1998ig}
\bibitem{Antoniadis:1998ig}
I.~Antoniadis, N.~Arkani-Hamed, S.~Dimopoulos and G.~R.~Dvali,
%``New dimensions at a millimeter to a Fermi and superstrings at a TeV,''
Phys.\ Lett.\ B {\bf 436}, 257 (1998)
[arXiv:hep-ph/9804398].
%%CITATION = HEP-PH 9804398;%%

%\cite{Arkani-Hamed:1998nn}
\bibitem{Arkani-Hamed:1998nn}
N.~Arkani-Hamed, S.~Dimopoulos and G.~R.~Dvali,
%``Phenomenology, astrophysics and cosmology of theories with  sub-millimeter
%dimensions and TeV scale quantum gravity,''
Phys.\ Rev.\ D {\bf 59}, 086004 (1999)
[arXiv:hep-ph/9807344].
 %%CITATION = HEP-PH 9807344;%%


%\cite{Randall:1999ee}
\bibitem{Randall:1999ee}
L.~Randall and R.~Sundrum,
%``A large mass hierarchy from a small extra dimension,''
Phys.\ Rev.\ Lett.\  {\bf 83}, 3370 (1999) [arXiv:hep-ph/9905221].
%%CITATION = HEP-PH 9905221;%%

%\cite{Randall:1999vf}
\bibitem{Randall:1999vf}
L.~Randall and R.~Sundrum,
%``An alternative to compactification,''
Phys.\ Rev.\ Lett.\  {\bf 83}, 4690 (1999)
[arXiv:hep-th/9906064].
%%CITATION = HEP-TH 9906064;%%

%\cite{Appelquist:2000nn}
\bibitem{Appelquist:2000nn}
T.~Appelquist, H.~C.~Cheng and B.~A.~Dobrescu,
%``Bounds on universal extra dimensions,''
Phys.\ Rev.\ D {\bf 64}, 035002 (2001)
[arXiv:hep-ph/0012100].
 %%CITATION = HEP-PH 0012100;%%

%\cite{Cheng:2002ab}
\bibitem{Cheng:2002ab}
H.~C.~Cheng, K.~T.~Matchev and M.~Schmaltz,
%``Bosonic supersymmetry? Getting fooled at the LHC,''
Phys.\ Rev.\ D {\bf 66}, 056006 (2002)
[arXiv:hep-ph/0205314].
%%CITATION = HEP-PH 0205314;%%

%\cite{Arkani-Hamed:2005px}
\bibitem{Arkani-Hamed:2005px}
N.~Arkani-Hamed, G.~L.~Kane, J.~Thaler and L.~T.~Wang,
%``Supersymmetry and the LHC inverse problem,''
arXiv:hep-ph/0512190.
%%CITATION = HEP-PH 0512190;%%

%\cite{ist}
\bibitem{ist}
J. P\"oschel and E. Trubowitz, ``Inverse Spectral Theory,'' Pure and Applied Mathematics,
Volume 130, Academic Press, 1987.

%\cite{Davoudiasl:1999tf}
\bibitem{Davoudiasl:1999tf}
H.~Davoudiasl, J.~L.~Hewett and T.~G.~Rizzo,
%``Bulk gauge fields in the Randall-Sundrum model,''
Phys.\ Lett.\ B {\bf 473}, 43 (2000)
[arXiv:hep-ph/9911262].
%%CITATION = HEP-PH 9911262;%%

%\cite{Muck:2001yv}
\bibitem{Muck:2001yv}
A.~Muck, A.~Pilaftsis and R.~Ruckl,
% ``Minimal higher-dimensional extensions of the standard model and  %electroweak observables,''
Phys.\ Rev.\ D {\bf 65}, 085037 (2002)
 [arXiv:hep-ph/0110391].
  %%CITATION = HEP-PH 0110391;


%\cite{Csaki:2003dt}
\bibitem{Csaki:2003dt}
C.~Csaki, C.~Grojean, H.~Murayama, L.~Pilo and J.~Terning,
%``Gauge theories on an interval: Unitarity without a Higgs,''
Phys.\ Rev.\ D {\bf 69}, 055006 (2004)
[arXiv:hep-ph/0305237].
 %%CITATION = HEP-PH 0305237;%%



%\cite{Georgi:2000ks}
\bibitem{Georgi:2000ks}
H.~Georgi, A.~K.~Grant and G.~Hailu,
%``Brane couplings from bulk loops,''
Phys.\ Lett.\ B {\bf 506}, 207 (2001)
 [arXiv:hep-ph/0012379].
 %%CITATION = HEP-PH 0012379;%%

%\cite{Cheng:2002iz}
\bibitem{Cheng:2002iz}
H.~C.~Cheng, K.~T.~Matchev and M.~Schmaltz,
%``Radiative corrections to Kaluza-Klein masses,''
Phys.\ Rev.\ D {\bf 66}, 036005 (2002)
[arXiv:hep-ph/0204342].
%%CITATION = HEP-PH 0204342;%%

%\cite{Cheng:2003ju}
\bibitem{Cheng:2003ju}
H.~C.~Cheng and I.~Low,
%``TeV symmetry and the little hierarchy problem,''
JHEP {\bf 0309}, 051 (2003)
[arXiv:hep-ph/0308199].
%%CITATION = HEP-PH 0308199;%%

%\cite{Cheng:2004yc,Low:2004xc}
\bibitem{Cheng:2004yc}
H.~C.~Cheng and I.~Low,
%``Little hierarchy, little Higgses, and a little symmetry,''
JHEP {\bf 0408}, 061 (2004)
  [arXiv:hep-ph/0405243].
%%CITATION = HEP-PH 0405243;%%

%\cite{Low:2004xc}
\bibitem{Low:2004xc}
I.~Low,
%``T parity and the littlest Higgs,''
JHEP {\bf 0410}, 067 (2004)
[arXiv:hep-ph/0409025].
%%CITATION = HEP-PH 0409025;%%



%\cite{Pomarol:1999ad}
\bibitem{Pomarol:1999ad}
A.~Pomarol,
%``Gauge bosons in a five-dimensional theory with localized gravity,''
Phys.\ Lett.\ B {\bf 486}, 153 (2000)
[arXiv:hep-ph/9911294].
%%CITATION = HEP-PH 9911294;%%

%\cite{Gherghetta:2000qt}
\bibitem{Gherghetta:2000qt}
T.~Gherghetta and A.~Pomarol,
%``Bulk fields and supersymmetry in a slice of AdS,''
Nucl.\ Phys.\ B {\bf 586}, 141 (2000)
[arXiv:hep-ph/0003129].
%%CITATION = HEP-PH 0003129;%%

%\cite{Arkani-Hamed:2001ca}
\bibitem{Arkani-Hamed:2001ca}
N.~Arkani-Hamed, A.~G.~Cohen and H.~Georgi,
%``(De)constructing dimensions,''
Phys.\ Rev.\ Lett.\  {\bf 86}, 4757 (2001)
[arXiv:hep-th/0104005].
%%CITATION = HEP-TH 0104005;%%

%\cite{Hill:2000mu}
\bibitem{Hill:2000mu}
C.~T.~Hill, S.~Pokorski and J.~Wang,
%``Gauge invariant effective Lagrangian for Kaluza-Klein modes,''
Phys.\ Rev.\ D {\bf 64}, 105005 (2001)
[arXiv:hep-th/0104035].
%%CITATION = HEP-TH 0104035;%%

%\cite{low}
\bibitem{low}
I.~Low and J.~Terning, work in progress.


%\cite{Rabadan:2002wb}
\bibitem{Rabadan:2002wb}
R.~Rabadan and G.~Shiu,
%``(Re)constructing dimensions,''
JHEP {\bf 0305}, 045 (2003)
[arXiv:hep-th/0212144].
%%CITATION = HEP-TH 0212144;%%


\end{thebibliography}
\end{document}